\documentclass[aps,prd,amsmath,amssymb,preprintnumbers,12pt,twoside, a4paper]{article}

\usepackage[dvipdfmx]{graphicx}
\usepackage{amsmath,amssymb,graphicx,epsfig}

\usepackage{color}

\newcommand{\re}{\text{Re}}

\usepackage{subfigure}
\usepackage{fancyhdr}

\input{epsf.sty}  
\begin{document}

\thispagestyle{empty}

\begin{flushright}
CERN-TH-2016-178\\
\end{flushright}

\begin{center}
{\Large{\textbf{Inflation from Supergravity with Gauged R-symmetry in de Sitter Vacuum}}}
\\
\medskip
\vspace{1cm}
\textbf{
I.~Antoniadis$^{\,a,b,}$\footnote{antoniadis@itp.unibe.ch}, 
A.~Chatrabhuti$^{c,}$\footnote{dma3ac2@gmail.com}, 
H.~Isono$^{c,}$\footnote{hiroshi.isono81@gmail.com}, 
R.~Knoops$^{\, d,e,}$\footnote{rob.knoops@cern.ch}
}
\bigskip

$^a$ {\small LPTHE, UMR CNRS 7589
Sorbonne Universit\'es, UPMC Paris 6,\\ 4 Place Jussieu, 75005 Paris, France}

$^b$ {\small Albert Einstein Center, Institute for Theoretical Physics,
University of Bern,\\ Sidlerstrasse 5, CH-3012 Bern, Switzerland }

$^c$ {\small Department of Physics, Faculty of Science, Chulalongkorn University,
\\Phayathai Road, Pathumwan, Bangkok 10330 , Thailand }

$^d$ {\small Section de Math\'ematiques, Universit\'e de Gen\`eve, CH-1211 Gen\`eve, Switzerland}

$^e$ {\small Instituut voor Theoretische Fysica, KU Leuven,
Celestijnenlaan 200D,\\ B-3001 Leuven, Belgium}

\end{center}

\vspace{1cm}
\begin{abstract}

We study the cosmology of a recent model of supersymmetry breaking, in the presence of a tuneable positive cosmological constant, based on a gauged shift symmetry of a string modulus that can be identified with the string dilaton. The minimal spectrum of the `hidden' supersymmetry breaking sector consists then of a vector multiplet that gauges the shift symmetry of the dilaton multiplet and when coupled to the MSSM leads to a distinct low energy phenomenology depending on one parameter. Here we study the question if this model can also lead to inflation by identifying the dilaton with the inflaton. We find that this is possible if the K\"ahler potential is modified by a term that has the form of NS5-brane instantons, leading to an appropriate inflationary plateau around the maximum of the scalar potential, depending on two extra parameters. This model is consistent with present cosmological observations without modifying the low energy particle phenomenology associated to the minimum of the scalar potential.

\end{abstract}
\newpage

\section{Introduction} 

A fundamental theory of Nature, such as string theory, should be able to describe at the same time particle physics and cosmology, which are phenomena that involve very different scales from the microscopic four-dimensional (4d) quantum gravity length of $10^{-33}$ cm to large macroscopic distances of the size of the observable Universe $\sim\! 10^{28}$ cm, spanned a region of about 60 orders of magnitude. In particular, besides the 4d Planck mass, there are  three very different scales with very different physics corresponding to the electroweak, dark energy and inflation. These scales might be related via the scale of the underlying fundamental theory, such as string theory, or they might be independent in the sense that their origin could be based on different and independent dynamics. 

In this work, we make an attempt towards this direction by connecting the scale of inflation with the electroweak and supersymmetry breaking scales within the same effective field theory, that at the same time allows the existence of an infinitesimally small (tuneable) positive cosmological constant describing the present dark energy of the universe. To this end, we use a simple model of supersymmetry breaking in a tuneable metastable de Sitter vacuum, independent of the scale of supersymmetry breaking that can be in the TeV region~\cite{Villadoro:2005yq}-\cite{Antoniadis:2015mna}. The model is based on a shift symmetry of a string modulus $S$ (that we identify with the string dilaton) along its imaginary (axionic) component, which is gauged by a $U(1)$ vector multiplet. The latter can be for instance a linear combination of the ordinary Baryon and Lepton number, containing the matter parity which guarantees a dark matter candidate~\cite{Antoniadis:2015adn}. 

The superpotential $W$ is completely fixed to a simple exponential ($W=ae^{bS}$, in a K\"ahler basis where the shift symmetry is an R-symmetry) depending on two parameters, while a third parameter arises from the $U(1)$ gauge coupling. On the other hand, the K\"ahler potential $\cal K$ is an arbitrary function of $S+{\bar S}$. However, since one is interested in a vacuum where the string coupling is weak (and therefore $S$ large), in order to study the supersymmetry phenomenology, one can restrict $\cal K$ to its tree-level logarithmic form, ${\cal K}=-p\log(S+{\bar S})$ with $p=1$ or 2, depending on whether $S$ is associated to only the D9 or to the D9 and D5 gauge couplings in type I string theory~\cite{Antoniadis:1996vw}. 

This model has the necessary ingredients to be obtained as a remnant of moduli stabilisation within the framework of internal magnetic fluxes in type I string theory, turned on along the compact directions for several abelian factors of the gauge group. All geometric moduli can in principle be fixed in a supersymmetric way, while the shift symmetry is associated to the 4d axion and its gauging is a consequence of anomaly cancellation~\cite{Antoniadis:2004pp,Antoniadis:2008uk}.

The resulting scalar potential has a metastable minimum with a tuneable positive vacuum energy that can be made infinitesimally small and one is left with one free parameter which fixes the scale of supersymmetry breaking. This is due to a tuning between the D- and the F-term contributions to supersymmetry breaking that can have opposite signs in supergravity. Coupling this model to the observable sector (MSSM) is straightforward but requires the introduction of an additional parameter in order to address the problem of anomaly cancellation or of tachyonic scalar masses for $U(1)$ neutral matter fields~\cite{Ghilencea,Antoniadis:2015mna}.

The main question we address in this work is whether the same scalar potential can provide inflation with the dilaton playing also the role of the inflaton at an earlier stage of the universe evolution. We show that this is possible if one modifies the K\"ahler potential by a correction that plays no role around the minimum, but creates an appropriate plateau around the maximum. In general, the K\"ahler potential receives perturbative and non-perturbative corrections that vanish in the weak coupling limit. After analysing all such corrections, we find that only those that have the form of (Neveu-Schwarz) NS5-brane instantons can lead to an inflationary period compatible with cosmological observations. The scale of inflation turns out then to be of the order of low energy supersymmetry breaking, in the TeV region. On the other hand, the predicted tensor-to-scalar ratio is too small to be observed. 

An interesting property of this model is that the inflaton is a component of the goldstino superpartner and shares some of the features of the models proposed in Ref.~\cite{AlvarezGaume:2010rt}\footnote{
For other models of sgoldstino inflation, see for example~\cite{Dine:2011ws}. However, the main difference with these models comes from the D-term contribution to sumersymmetry breaking. 
}, 
although it does not belong to the same class, since there is a D-term contribution to supersymmetry breaking. As a result, the gravitino mass does not satisfy the property to be much lower than the supersymmetry breaking scale.

The outline of the paper is the following. In Section~2, we give a brief review of the model and compute the slow-roll parameters of the scalar potential to show that without modifying the K\"ahler potential, it does not give rise to inflation (subsection 2.1). We then present its general extension, parametrising appropriately the quantum corrections to the K\"ahler potential (subsection 2.2). In Section~3, we perform the analysis for a correction that has the form of NS5-brane instantons (depending on two additional parameters) for both $p=2$ (subsection 3.1) and $p=1$ (subsection 3.2) cases. In particular, we compute the corresponding slow-roll parameters and fix the two parameters of the correction, so that our model reproduces the spectral index and amplitude of density fluctuations in agreement with the cosmological data of Planck '15. We also extract the predictions for the inflation scale, the number of e-foldings and the tensor-to-scalar ratio. In Section~4, we study the impact of the correction to the K\"ahler potential for the low energy superparticle spectrum that we find to be of order of 10\%.
Finally, Section~5 contains our concluding remarks.


\section{The model}
\subsection{Revision of the model}

In \cite{Villadoro:2005yq,Knoops} a $\mathcal N=1$ supergravity model was proposed based on the gauged shift symmetry of a single chiral multiplet.
For certain values of the parameters, the model allows for a tunably small and positive value for the cosmological constant. 
Its anomaly cancelation conditions are discussed in~\cite{Ghilencea}, while the resulting low energy spectrum is discussed in~\cite{Antoniadis:2015mna,Antoniadis:2015adn}.
In this section we recall the main properties of the model. We use the conventions of~\cite{Freedman:2012zz}.

The model consists of one chiral multiplet $S$, whose scalar component $s$ is invariant under a gauged shift symmetry\footnote{
For other models based on a (global) shift symmetry, see for example~\cite{Kawasaki:2000yn}. }
\begin{align} s \longrightarrow s - ic \theta, \label{shift} \end{align}
where $\theta$ is the gauge parameter, and $c$ is a constant. 
The K\"ahler potential, superpotential and gauge kinetic function are given by (in appropriate K\"ahler coordinates where the superpotential is constant) 
\begin{align}
 \mathcal K(s, \bar s) &= - \kappa^{-2} p \log(s + \bar s) + \kappa^{-2} b (s + \bar s), \notag \\
 W(s) &= \kappa^{-3} a, \notag \\
 f(s) &= \gamma + \beta s ,
\end{align}
where $\kappa^{-1} = m_p = 2.4 \times 10^{15}$ TeV is the inverse of the (reduced) Planck mass. 
The scalar potential is given by
\begin{align} \mathcal V &= \mathcal V_F + \mathcal V_D, \notag \\ 
\mathcal V_F &= e^{\kappa^2 \mathcal K} \left( - 3 \kappa^2 W \bar W  + \nabla_\alpha W g^{\alpha \bar \beta} \bar \nabla_{\bar \beta} \bar W \right),  \notag \\ 
\mathcal V_D &= \frac{1}{2} \left( \re f \right)^{-1 \ AB} \mathcal P_A \mathcal P_B, \label{scalarpot} 
\end{align}
where Greek indices $\alpha, \beta$  label the chiral multiplets in the theory, and capital Roman letters $A,B$ label the different gauge groups.
In eqs.~(\ref{scalarpot}), the K\"ahler covariant derivative of the superpotential is 
  \begin{align}\nabla_\alpha W = \partial_\alpha W(z) + \kappa^2 (\partial_\alpha \mathcal K) W(z),\end{align}
  and the moment maps $\mathcal P_A$ are given by
  \begin{align} \mathcal P_A = i(k_A^\alpha \partial_\alpha \mathcal K - r_A), \label{momentmap} \end{align}
where $k_A^\alpha$ are the Killing vectors, and $r_A$ is the Fayet-Iliopoulos contribution satisfying $W_\alpha k_A^\alpha = - \kappa^2 r_A W$.
In the above model $k^s = -ic$ is the Killing vector associated with the shift symmetry eq.~(\ref{shift}).
As a result, the scalar potential is given by (with $\phi = s + \bar s$)
\begin{align} \mathcal V &= \mathcal V_F + \mathcal V_D, \notag \\
\mathcal V_F &= \frac{ \kappa^{-4} |a|^2 }{\phi^p} e^{b\phi} \left( -3 + \frac{1}{p}  \left( b\phi - p \right)^2 \right) \notag \\
\mathcal V_D &=  \frac{\kappa^{-4} c^2}{\beta\phi + 2\gamma}  \left( b - \frac{p}{\phi} \right)^2 . \label{scalarpot_p}
 \end{align}
For $b > 0$, the potential always admits a supersymmetric AdS (anti-de Sitter) vacuum at $\langle \phi \rangle = b/p$, while for $b=0$ supersymmetry is broken in AdS space.
We therefore focus on $b<0$.
In this case, the potential admits a supersymmetry breaking dS (de Sitter) vacuum for $p < 3$.
For example, for $p=2$, $\beta =1$ and $\gamma = 0$,\footnote{
As far as the scalar potential is concerned, the parameter $\beta$ can indeed always be absorbed into other parameters of the model.
Although we assumed $\gamma = 0$, a very small $\gamma$ is allowed and in fact consistent with a vanishing cosmological constant.
However, the resulting change in the scalar potential can be neglected and we therefore take $\gamma=0$.}
the scalar potential eq.~(\ref{scalarpot_p}) reduces to
\begin{align} \mathcal V &= \frac{ \kappa^{-4} |a|^2 }{\phi^2} e^{b\phi} \left( -3 + \frac{1}{2}  \left( b\phi - 2 \right)^2 \right) +  \frac{\kappa^{-4} c^2}{\phi}  \left( b - \frac{2}{\phi} \right)^2 . \label{scalarpot_2}
 \end{align}
A vanishing cosmological constant can be found by tuning the parameters of the model. 
Solving the equations $\mathcal V(\phi_{ \text{min} }) = 0$ and $d \mathcal V( \phi_{\text{min} } ) / d\phi = 0$ gives
\begin{align}
 b \phi_{\text{min}} &= l_0 \approx -0.183268, \label{vacuum1}
\end{align}
where $l_0$ is the root of the polynomial $-x^5 + 7 x^4 - 10 x^3 -22 x^2 +40 x + 8$ close to $-0.18$, 
and
\begin{align}
 \frac{a^2}{b c} = \mathcal A(l_0) \approx -50.66 .\label{vacuum2}
\end{align}
In eq.~(\ref{vacuum2}), $\mathcal A(l_0)$ is given by
\begin{align}
 \mathcal A(l_0)
 &= \frac{e^{-l_0} }{l_0} \left( \frac{- 4 + 4 l_0  - l_0^2}{\frac{l_0}{2} - 2 l_0 -1} \right)  .
\end{align}
A non-zero cosmological constant $\Lambda$ can be found if 
\begin{align}  \frac{a^2}{bc^2}  = \mathcal A (l_0)+ \frac{\kappa^4 \Lambda}{b^3 c^2} \left( \frac{l_0^2 e^{-l_0}}{\frac{l_0^2}{2} - 2l_0 -1}  \right) .  \end{align}
The gravitino mass parameter is given by 
 \begin{align}  m_{3/2} = \kappa^2 e^{\kappa^2 \mathcal K/2} W = \frac{ \kappa^{-1} a b}{l_0} e^{l_0/2}  . \label{gravitinomass} \end{align}
The first relation (\ref{vacuum1}) fixes the VEV (Vacuum Expectation Value) of $\phi$ as a function of the parameter $b$. 
The field $\phi$ can be interpreted as the dilaton\footnote{
Note that the gauge kinetic function is given by $f(s) = s$ for $p=2$, such that we indeed have $\mathcal L/e  \ni -\frac{1}{4} \frac{ \phi_{ \text{min} } }{2} F_{\mu \nu} F^{\mu \nu}$.}
related to the string coupling constant $g_s$ by $\phi_{\text{min}} = 2/g_s$. 
A string coupling constant in the perturbative region, for example $\phi_{\text{min}} = 10$, can therefore be obtained for example by the choice $b = -0.01820$.
Since the parameters $a$ and $c$ are related by eq.~(\ref{vacuum2}), this leaves only one free parameter, which can be tuned to obtain a $O(10 \text{ TeV})$ gravitino mass.
For example, the parameter choice $c=0.61 \times 10^{-13}$ results in $m_{3/2} = 12.83 \text{ TeV}$.
However, we will show below that this model does not allow slow roll inflation.

The kinetic terms in the Lagrangian for the scalar $\phi$ are given by
\begin{align}
 \mathcal L_s / e &= -g_{s \bar s} \partial_\mu s \partial^\mu \bar s \notag \\
 & 
 = -\frac{p\kappa^{-2} }{4} \frac{1}{\phi^2} \partial_\mu \phi \partial^\mu \phi .
\end{align}
The canonically normalised field $\chi$ therefore satisfies $\chi = \kappa^{-1} \sqrt{\frac{p}{2}} \log \phi$.

\begin{figure}[h!]
    \centering
            \includegraphics[width=0.45\textwidth]{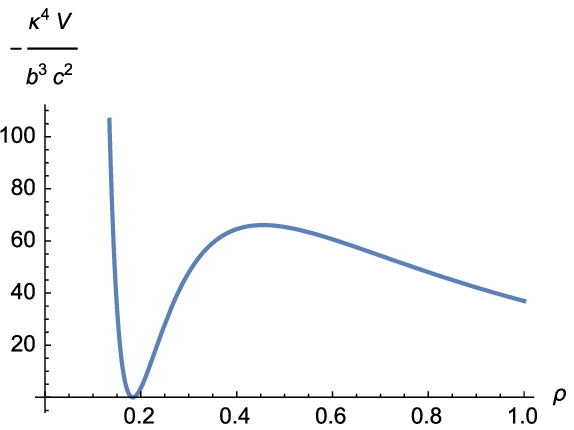}
              \includegraphics[width=0.45\textwidth]{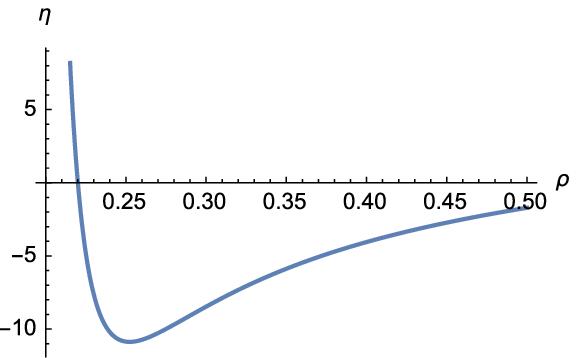}
     \caption{A plot of $ -\frac{\kappa^4 \mathcal V (\rho)}{b^3 c^2}$ as a function of $\rho = - b \phi$ (left),
and a plot of the slow roll parameter $\eta$ as a function of $\rho$ (right). The slow roll condition $|\eta| \ll 1$ is not satisfied for any value of the parameters $a,b,c$.}
\label{fig:xi0}
\end{figure}

The slow roll parameters are given by
\begin{align}
 \epsilon &= \frac{1}{2\kappa^2} \left( \frac{dV/d\chi}{V}\right)^2 = \frac{1}{2\kappa^2}\left[\frac{1}{V} \frac{dV}{d\phi} \left( \frac{d\chi}{d\phi} \right)^{-1} \right]^2, \notag \\
 \eta &= \frac{1}{\kappa^2} \frac{V''(\chi)}{V} =  \frac{1}{\kappa^2} \frac{1}{V} \left[ \frac{d^2V}{d\phi^2} \left( \frac{d\chi}{d\phi}\right)^{-2} - \frac{dV}{d\phi} \frac{d^2 \chi}{d\phi^2} \left(\frac{d \chi}{d\phi} \right)^{-3} \right],
 \label{slowroll_pars}
\end{align}
It can be shown that, when the conditions (\ref{vacuum1}) and (\ref{vacuum2}) are satisfied, the slow roll parameters depend only on $\rho = - b \phi$
\begin{align}
 \epsilon &= \frac{(\rho +2)^2 \left( \mathcal A(l_0) \rho  \left(\rho ^2+2 \rho -2\right)-2 e^{\rho } (\rho +6)\right)^2}{2 \left( \mathcal A(l_0)  \rho  \left(\rho ^2+4 \rho -2\right)-2 e^{\rho } (\rho +2)^2\right)^2} , \notag \\
\eta&= \frac{2 e^{\rho } \left(3 \rho ^2+32 \rho +60\right) - \mathcal A(l_0) \rho  \left(\rho ^4+5 \rho ^3+10 \rho ^2+2 \rho -16\right)}{2 e^{\rho } (\rho +2)^2-\mathcal A(l_0)  \rho  \left(\rho ^2+4 \rho -2\right)} .
\end{align}
Moreover, the scalar potential is given in terms of $\rho$
\begin{align}
  \frac{\kappa^4 \mathcal V (\rho)}{b^3 c^2} &= \frac{e^{-\rho } \left(\mathcal A(l_0) \rho  \left(\rho ^2+4 \rho -2\right)-2 e^{\rho } (\rho +2)^2\right)}{2 \rho ^3} ,
\end{align}
where $\mathcal A(l_0) \approx -50.66$ as in eq.~(\ref{vacuum2}).
 In Fig.~\ref{fig:xi0}, a plot is shown of  $ \frac{\kappa^4 \mathcal V (\rho)}{|b|^3 c^2}$ as a function of $\rho$. 
The minimum of the potential is at $\rho_\text{min} \approx 0.1832$ (see eq.~(\ref{vacuum1})), while the potential has a local maximum at $\rho_\text{max} \approx 0.4551$.
A plot of the slow roll parameter $\eta$ (also in Fig.~\ref{fig:xi0}) shows that $|\eta| \ll 1$ is not satisfied.
This result holds for any parameters $a,b,c$ satisfying eqs.~(\ref{vacuum1}) and (\ref{vacuum2}).

A similar analysis to the one above can be performed for $p=1$.
In contrast with the case $p=2$, vacua with a vanishing cosmological constant can also be found when $\beta = 0$ and $\gamma \neq 0$.
Note however that for $\beta \neq 0$ the Lagrangian contains a Green-Schwarz term 
\begin{align} \mathcal L_{GS} &=  \frac{\beta \ \text{Im}\left(s \right) }{8}  \ \epsilon^{\mu \nu \rho \sigma} F_{\mu \nu} F_{\rho \sigma},\label{Green-Schwarz} \end{align}
which is not gauge invariant. 
In principle, such a term can be present if the model contains additional fields, that are charged under the $U(1)$ gauge symmetry (see eq.~(\ref{shift})), 
in order to cancel their contribution to a possible cubic $U(1)^3$ anomaly via a Green-Schwarz mechanism.
However, it turns out (see for example~\cite{Antoniadis:2015adn}) that even in this case $\beta$ is very small.
We therefore take $\beta = 0$ for simplicity.\footnote{The Green-Schwarz term eq.~(\ref{Green-Schwarz}) is always present in the case $p=2$ above, 
since in this case one can only find vacua with a vanishing cosmological constant when $\beta > 0$.
 This is the reason why we focus on $p=1$ below.}
Moreover, as long as the scalar potential is concerned, $\gamma$ can be absorbed in other parameters of the model, and we therefore take $\gamma = 1 $. 

For $\gamma=1$ and $\beta=0$, vacua with a vanishing cosmological constant can be found if $b\phi_\text{min} =l_0 \approx  -0.233153$, and $\frac{a^2}{b c^2} = \mathcal A(l_0) \approx  -2.783259$.
However, a similar analysis as the one above shows that also in this case the slow roll condition $\eta \ll 1$ can not be satisfied.


\subsection{Extensions of the model that satisfy the slow roll conditions}

In the previous section we showed that the slow roll conditions can not be satisfied in the minimal versions of the model.
In this section we modify the above model by modifying the K\"ahler potential. 
While the superpotential is uniquely fixed (up to a K\"ahler transformation\footnote{ 
For example, by performing a K\"ahler transformation
\begin{align} \mathcal K(z ,\bar z) &\longrightarrow \mathcal K(z , \bar z) + J(z) + \bar J(\bar z),  \notag \\ W(z) &\longrightarrow e^{-\kappa^2 J(z)} W(z), \notag \end{align}
  with $J(z) = -\kappa^{-2}bs$, the linear contribution in the K\"ahler potential can be absorbed into the superpotential, which becomes of the form $W(s) = \kappa^{-3}a \exp{(bs)}$.
}), the K\"ahler potential admits corrections that can always be put in the form 
\begin{align}\label{Kgeneral}
  \mathcal K&= -p \kappa^{-2} \log \left(s + \bar s  + \frac{\xi}{b} F(s + \bar s) \right) + \kappa^{-2} b (s + \bar s),
\end{align}
while the superpotential, the gauge kinetic function and moment map are given by
\begin{align}\label{WfPgeneral}
W&= \kappa^{-3} a, \notag \\
 f(s) &=  \gamma + \beta s, \notag \\
 \mathcal P &= \kappa^{-2} c \left( b - p   \frac{ 1 + \frac{\xi}{b} F_s  }{s + \bar s + \frac{\xi}{b} F } \right), 
\end{align}
where $F_s = \partial_s F(s + \bar s)$.
The scalar potential is given by ($\phi = s + \bar s$)
\begin{align}
 \mathcal V &= \mathcal V_F + \mathcal V_D , \notag \\ 
 \mathcal V_F &=  \kappa^{-4} \frac{|a|^2 e^{b\phi} }{(  \phi + \frac{\xi}{b} F)^p} 
\left[ -3 - \frac{1}{p} \frac{ \left( b \left( b \phi + \xi F \right) - p (b +  \xi F_\phi ) \right)^2}{ \xi  F_{\phi \phi} ( b \phi + \xi F)  - (b + \xi F_\phi)^2} \right], \notag \\
 \mathcal V_D &= \kappa^{-4}  \frac{b^2 c^2}{2 \gamma + \beta \phi} \left[ 1 - p \frac{ 1 + \frac{\xi}{b} F_\phi }{b \phi + \xi F} \right]^2.
\end{align}
As was discussed above, we take $\gamma = 1, \beta = 0$ for $p=1$ and $\gamma=0, \beta =1$ for $p=2$.

Identifying $\text{Re}(s)$ with the inverse string coupling, the function $F$ may contain perturbative contributions that can be expressed as power series of $1/(s+\bar{s})$, as well as non-perturbative corrections which are exponentially suppressed in the weak coupling limit. The later can be either of the form $e^{-\delta(s+\bar{s})}$ for $\delta>0$ in the case of D-brane instantons, or of the form $e^{-\delta(s+\bar{s})^2}$ in the case of (Neveu-Schwarz) NS5-brane instantons (since the closed string coupling is the square of the open string coupling). We have considered a generic contribution of these three different types of corrections and we found that only the last type of contributions can lead to an inflationary plateau providing sufficient inflation. 
The other corrections can be present but do not modify the main properties of the model (as long as weak coupling description holds).
In the following section, we analyse in detailed a function $F$ describing a generic NS5-brane instanton correction to the K\"ahler potential.

\section{Slow-roll Inflation}
\subsection{p=2 case}
\begin{figure*}
\begin{center}
  \includegraphics[width=0.8\linewidth]{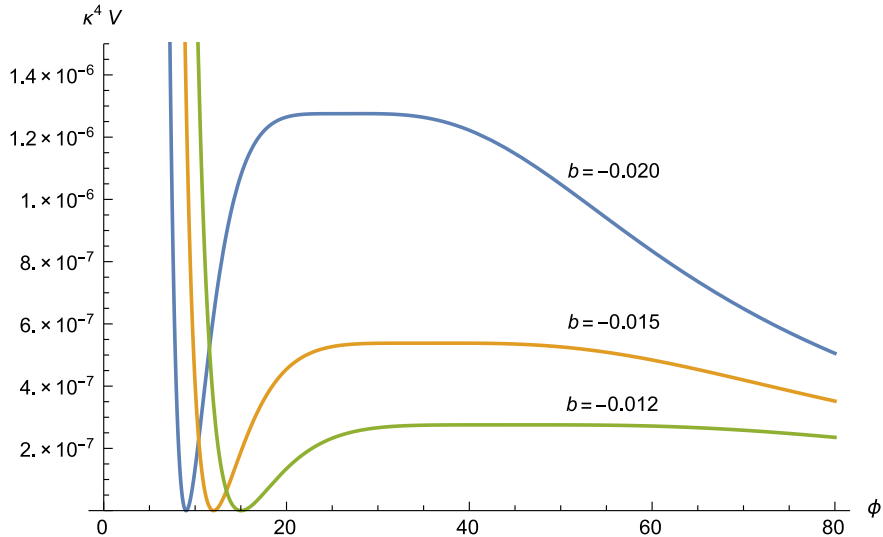}
 \caption{A plot of the scalar potential for $p=2$, with $b=-0.020$, $b=-0.015$ and $b = -0.012$.   Note that we choose the parameters $\alpha$ and $\xi$ as in eq. (\ref{fix_ax}) with $c = 0.06$.}
 \label{Potential_b_change}
\end{center}
\end{figure*}

\begin{figure*}
\begin{center}
 \includegraphics[width=0.8\linewidth]{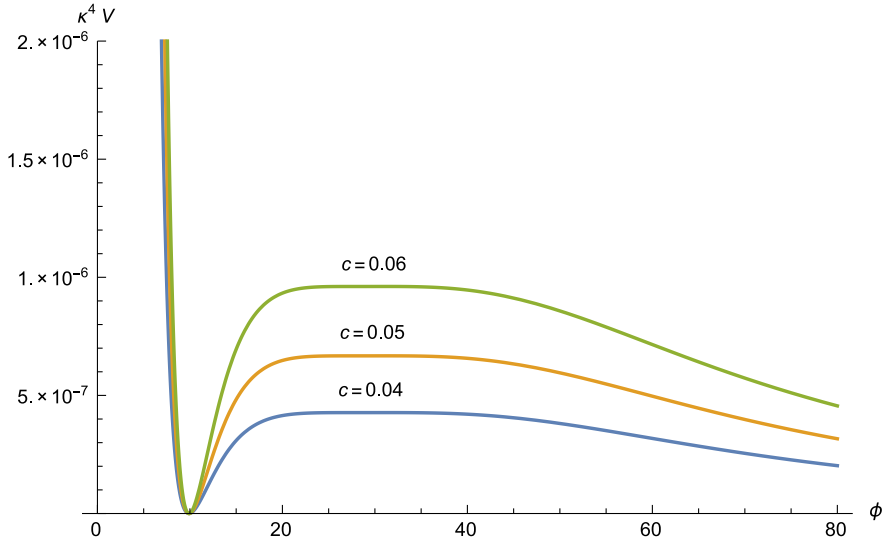}
 \caption{A plot of the scalar potential for $p=2$, with $c=0.04$, $c=0.05$ and $c = 0.06$.   Note that we choose the parameters $\alpha$ and $\xi$ as in eq. (\ref{fix_ax}) with $b = -0.0182$.}
 \label{Potential_c_change}
\end{center}
\end{figure*}

We now consider the case with
\begin{equation}
F(\phi) = \exp( \alpha b^2 \phi^2) ,
\end{equation}
where $b<0$ and $\alpha<0$ .  $F(\phi)$ vanishes  asymptotically at large $\phi$.   In this case, we obtain
\begin{equation}
\mathcal V_D = \frac{\kappa^{-4}b^3 c^2}{b\phi}\left[ \frac{b \phi - 2 + \xi e^{\alpha b^2 \phi^2}(1-4 \alpha b \phi )}{b \phi + \xi e^{\alpha b^2 \phi^2}}\right]^2,
\end{equation}
and
\begin{equation}
\mathcal V_F = -\frac{\kappa^{-4}|a|^2 b^2 e^{b \phi}}{2  \left(\xi  e^{\alpha  b^2 \phi^2}+b\phi\right)^2} \left[\frac{\left( b\phi+\xi  e^{\alpha  b^2 \phi^2 }(1-4 \alpha b \phi) -2\right)^2}{2 \alpha   \xi  e^{\alpha  b^2 \phi^2} \left(2 \alpha  b^3 \phi^3+\xi  e^{\alpha  b^2 \phi^2}-b\phi\right)-1}+6\right]. 
\end{equation}
There are four parameters in this model namely $\alpha$, $\xi$, $b$ and $c$.  The first two parameters $\alpha$ and $\xi$ control the shape of the potential.  There are some regions in the parameter space of $\alpha$ and $\xi$ that the potential satisfies the slow-roll conditions i.e. $\epsilon \ll 1$ and $|\eta| \ll 1$.  In order to obtain the potential with flat plateau shape which is suitable for inflation and in agreement with Planck '15 data, we choose \footnote{Large amount of significant digits for $\alpha$, $\xi$, $\lambda$ and $\phi_{int}$  are necessary to make our results reproducible.  They are necessary to tune the cosmological constant at the minimum and to create an inflationary plateau around the (local) maximum.
}
\begin{equation}
\alpha = -4.841115384560439  \text{ and } \xi = 0.025350051999999998. 
\label{fix_ax}
\end{equation}
Note that in the case of $\xi =0$ and $b < 0$, we can find the Minkowski minimum by solving the equations $\mathcal V(\phi_{min})=0$ and $d \mathcal V(\phi_{min})/d\phi =0$, where $\phi_{min} = s_{min}+\bar{s}_{min}$ is the value of $\phi$ at the minimum of the potential.  In the case of $\xi \neq 0$, we can not solve the equations analytically and the relations  (\ref{vacuum1}), (\ref{vacuum2}) are not valid. We can always assume that they are modified into 
\begin{equation}
 b \phi_{min} = l(\xi,\alpha) ~\text{  and  }~ \frac{a^2}{bc^2} =  -50.66016761885055 \times \lambda(\xi,\alpha,\Lambda)^2,
\label{Minkowski_condition_p=2}
\end{equation}
where $\lambda$ takes positive values and satisfies $|\lambda - 1| \ll 1$.  For any given value of parameters $\xi$,  $\alpha$ and the cosmological constant $\Lambda$, one can numerically fix the value of $l$ and $\lambda$.
By fine-tuning the cosmological constant $\Lambda$ to be very close to zero, we can numerically solve the equations $\mathcal V=0$ and $d\mathcal V/d\phi = 0$ for the value of $l$ and $\lambda$ in (\ref{Minkowski_condition_p=2}) as:
\begin{eqnarray}
l &\approx& -0.180386,\\
\lambda&=& 1.0172553241157374, 
\label{fix_Minkowski}
\end{eqnarray}
From eq. (\ref{Minkowski_condition_p=2}), we can see that the third parameter, $b$, controls the vacuum expectation value $\phi_{min}$.   This can be shown in Fig.~\ref{Potential_b_change} where we compare the scalar potential for different values of $b$.  Motivated by string theory, we have the identification $\phi \sim 1/g_s$ .  We can choose the value of the parameter $b$ such that $\phi_{min}$ is of the order of 10 to make sure that we are in the perturbative regime in $g_s$.  The last parameter, $c$, controls the overall scale of the potential but does not change its minimum and its shape.  This can be shown in Fig.~\ref{Potential_c_change}.   In the following, we will fix $b$ and $c$ by using the cosmological data.

In order to compare the predictions of our models with Planck '15 data, we choose the following boundary conditions: 
\begin{eqnarray}
\phi_{int} =  27.3204582176, \label{phi_int} \\
\phi_{end} = 22.6843813287. \label{phi_end}
\end{eqnarray}
The initial conditions are chosen very near the maximum on the (left) side, so that the field rolls down towards the electroweak minimum. 
Any initial condition on the right of the maximum may produce also inflation, but the field will roll towards the SUSY vacuum at infinity.
The results are therefore very sensitive to the initial conditions (eqs.~(\ref{phi_int}), (\ref{phi_end})) of the inflaton field.

The slow roll parameters are given as in equation (\ref{slowroll_pars}).  The total number of e-folds $N$ can be determined by
\begin{equation}
N= \kappa^2 \int_{\chi_{end}}^{\chi_{int}} \frac{\mathcal V}{\partial_{\chi} \mathcal V} d\chi =  \kappa^2 \int_{\phi_{end}}^{\phi_{int}} \frac{\mathcal V}{\partial_\phi\mathcal V}\left(\frac{d\chi}{d\phi} \right)^2 d\phi.
\end{equation}
Note that we choose $|\eta(\chi_{end})| = 1$.  We can compare the theoretical predictions of our model to the experimental results via the power spectrum of scalar perturbations of the CMB, namely the amplitude $A_s$ and tilt $n_s$, and the relative strength of tensor perturbations, i.e. the tensor-to-scalar ratio $r$. In terms of slow roll parameters, these are given by
\begin{eqnarray}
A_s &=& \frac{\kappa^4 \mathcal V_*}{24\pi^2 \epsilon_*},\\
n_s &=& 1 + 2\eta_* - 6\epsilon_*,\\
r &=& 16 \epsilon_*,
\end{eqnarray}
where all parameters are evaluated at the field value $\chi_{int}$.  

\begin{figure*}
\begin{center}
 \includegraphics[width=0.48\linewidth]{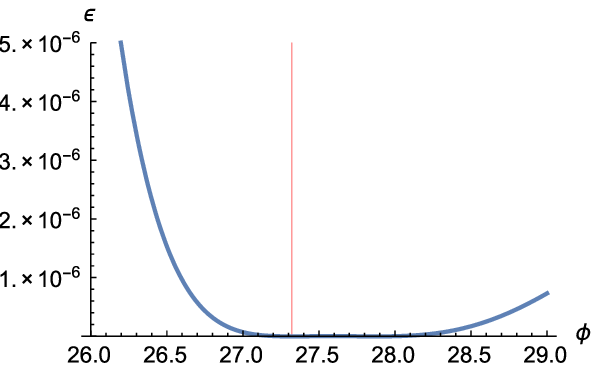}
 \includegraphics[width=0.48\linewidth]{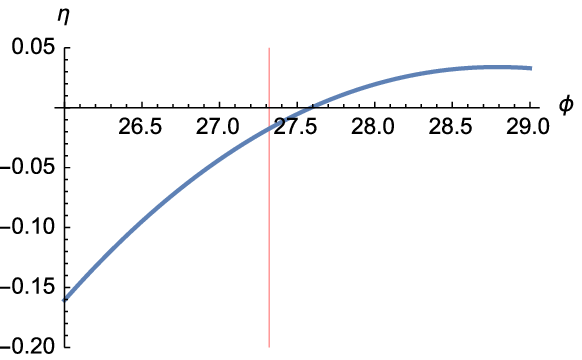}
 \caption{A plot of $\epsilon$  and $\eta$ versus $\phi$ for $p=2$, with $b=-0.0182$, $c=0.61 \times 10^{-13}$, $\xi = -0.1392$, where the parameters $\alpha$ and $\xi$ are given in eq (\ref{fix_ax}). The vertical (red) line indicates the initial value of $\phi$.}
 \label{Epsilon_Eta}
\end{center}
\end{figure*}

\begin{figure*}
\begin{center}
 \includegraphics[width=1.0\linewidth]{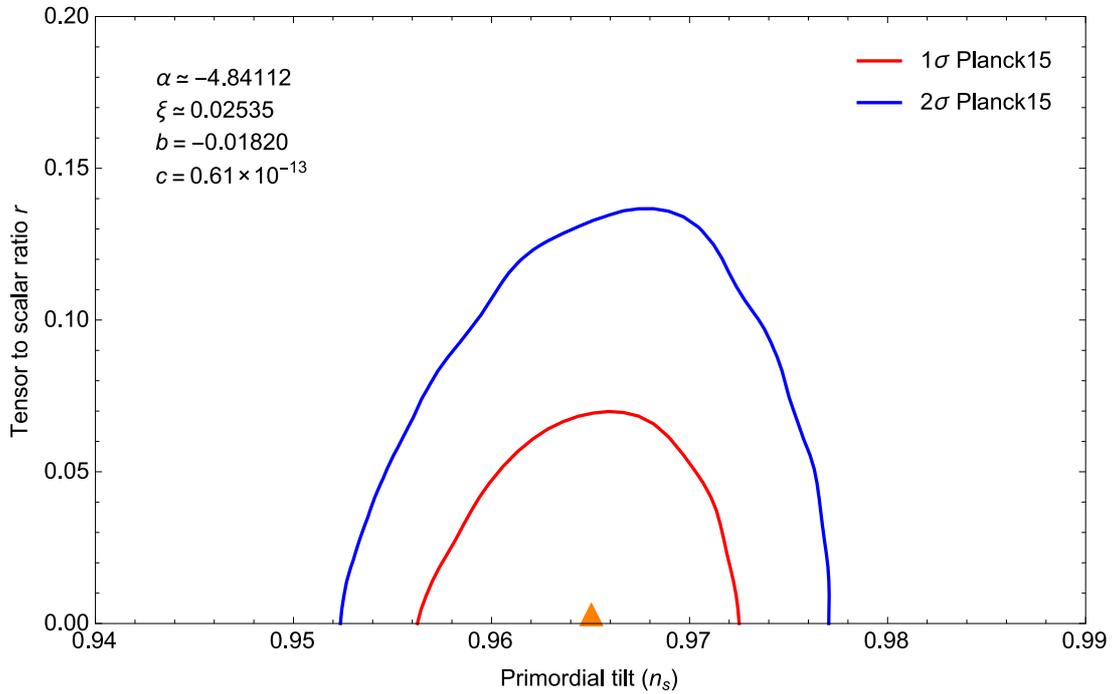}
 \caption{We plot the theoretical predictions  for the case $p=2$, shown in Table \ref{prediction}, in the $n_s$ - $r$ plane together with the Planck '15 results for TT, TE, EE, + lowP and assuming $\Lambda$CDM + r \cite{Ade:2015lrj}.}
 \label{n_r_Plot_I}
\end{center}
\end{figure*}

In order to satisfy Planck '15 data, we choose the parameters $b = -0.0182 $, $c=0.61 \times 10^{-13}$.  The slow-roll parameters $\epsilon$ and $\eta$ during the inflation are shown in Fig.~\ref{Epsilon_Eta}.   The value of the slow-roll parameters at the beginning of inflation are
\begin{equation}
\epsilon(\phi_{int}) \simeq 1.86 \times 10^{-24} \text{ and } \eta(\phi_{int}) \simeq -1.74 \times 10^{-2}.
\end{equation}
The total number of e-folds $N$, the scalar power spectrum amplitude $A_s$, the spectral index of curvature perturbation $n_s$ and the tensor-to-scalar ratio $r$ are calculated and   summarised in Table~\ref{prediction}, in agreement with Planck '15 data~\cite{Ade:2015lrj}.   Fig.~\ref{n_r_Plot_I} shows that our predictions for $n_s$ and $r$ are within 1$\sigma$ C.L. of Planck '15 contours with the total number of e-folds $N \approx 1075$.  Note that $N$ is the total number of e-folds from $\phi_{int}$ to $\phi_{end}$.  However the number of e-folds associated with the CMB observation corresponds to a period between the time of horizon crossing and the end of inflation, which is much smaller than $1075$.   According to general formula in \cite{Ade:2015lrj}, the number of e-folds between the horizon crossing and the end of inflation is roughly estimated to be around 50-60.
\begin{table}
  \centering 
  \begin{tabular}{|c|c|c|}
\hline
 $n_s$ & $r$ & $A_s$  \\
\hline
 0.965 & $2.969 \times 10^{-23} $ & $2.259  \times 10^{-9} $  \\
\hline
\end{tabular}
  \caption{The theoretical predictions for $p=2$, with $b =-0.0182$ and $c = 0.61\times 10^{-13}$, where $\alpha$ and $\xi$ are given in eq.~(\ref{fix_ax}). }
  \label{prediction}
\end{table}

We would like to remark that the parameter $c$ also controls the gravitino mass at the minimum of the potential around $O(10)$ TeV.  Indeed, the gravitino mass is written as 
\begin{equation}
m_{3/2} = \kappa^2 e^{\kappa^2\mathcal{K}/2 } W  = \frac{1}{\kappa}\left(\frac{a b e^{b \phi/2}}{b \phi + \xi F(\phi)} \right). \label{gravitino_mass}
\end{equation}
For $b=-0.0182$, we get $\phi_{\min} \approx 9.91134$ and the gravitino mass at the minimum of the potential 
\begin{equation}
\left<m_{3/2}\right> \approx 14.98  \text{  TeV}.
\end{equation}
The Hubble parameter during inflation (evaluated at $\phi_* = \phi_{int}$) is
\begin{equation}
H_* = \kappa \sqrt{\mathcal V_* / 3} = 1.38 \text{  TeV}.
\end{equation}
This shows that our predicted scale for inflation is of the order of TeV. The mass of gravitino during the inflation $m^*_{3/2} =  4.15$ TeV is higher than the inflation scale, 
and the gauge boson mass is $M_{A_\mu}^* = 3.12$~TeV.\footnote{The gauge boson mass is given by $m_{A_\mu} = \sqrt{2 g_{s\bar{s}} c^2 / \text{Re}(s)}$.}  In fact, the gauge boson acquires a mass due to a Stueckelberg mechanism by eating the imaginary component of $s$, where its mass at the minimum of the potential is given by
\begin{align} \langle M_{A_\mu} \rangle = 15.48 \text{ TeV}.\end{align}
As a result, the model essentially contains only one scalar field $\text{Re}(s)$, which is the inflaton.
This is in contrast with other supersymmetric models of inflation, which usually contain at least two real scalars~\cite{Baumann:2011nk}.\footnote{
This is because a chiral multiplet contains a complex scalar.}

\subsection{p=1 case} \label{sec:p=1}
In this case, we obtain
\begin{equation}
\mathcal V_D = \frac{\kappa^{-4}b^2 c^2}{2}\left[ \frac{b \phi - 1 + \xi e^{\alpha b^2 \phi^2}(1-2 \alpha b \phi )}{b \phi + \xi e^{\alpha b^2 \phi^2}}\right]^2,
\end{equation}
and
\begin{equation}
\mathcal V_F = -\frac{\kappa^{-4}|a|^2 b e^{b \phi}}{  \xi  e^{\alpha  b^2 \phi^2}+b\phi} \left[\frac{\left( b\phi+\xi  e^{\alpha  b^2 \phi^2 }(1-2 \alpha b \phi) -1\right)^2}{2 \alpha   \xi  e^{\alpha  b^2 \phi^2} \left(2 \alpha  b^3 \phi^3+\xi  e^{\alpha  b^2 \phi^2}-b\phi\right)-1}+3\right]. 
\end{equation}  
The potential has similar properties with the $p=2$ case although it may give different phenomenological results at low energy.   For $\xi =0$ and $b < 0$, the Minkowski minimum satisfies the following relations \cite{Knoops}
\begin{equation}
 b \phi_{min} = l_0 \approx -0.233153 ~\text{  and  }~ \frac{bc^2}{a^2} = \mathcal{A}(l_0) \approx - 0.35929085159984514. \label{vacuum_p=1}
\end{equation}
However, similar to the previous case, the above relations are not valid when $\xi \neq 0$ and we assume that they are modified into 
\begin{equation}
 b \phi_{min} = l(\xi,\alpha) ~\text{  and  }~ \frac{bc^2}{a^2} =  - 0.35929085159984514 \times \lambda(\xi,\alpha,\Lambda)^{-2}.
\label{Minkowski_condition_p=1}
\end{equation}
By choosing $\alpha = -0.781$ and $\xi = 0.3023$ and tuning the cosmological constant $\Lambda$ to be very close to zero, we can numerically fix $l = -0.562536$ and $\lambda = 1.2937645$ for this case.  The gravitino mass for $p=1$ case can be written as
\begin{equation}
m_{3/2} = \kappa^2 e^{\kappa^2\mathcal{K}/2 } W = \frac{1}{\kappa}\left( \frac{a \sqrt{b} e^{b \phi/2}}{\sqrt{b \phi+ \xi F(\phi)}} \right)\, .
\end{equation}
By choosing the parameters $b=-0.0234$, $c=1 \times 10^{-13}$, the gravitino mass at the minimum of the potential is
\begin{align} \langle m_{3/2} \rangle = 18.36 \text{ TeV}. \end{align}
 with $\phi_{\min} \approx 21.53$, and
\begin{align} \langle M_{A_\mu} \rangle = 36.18 \text{ TeV}.\end{align} 
  By choosing appropriate boundary conditions, we find
\begin{eqnarray}
\phi_{int} &=&  64.5324301784,\\
\phi_{end} &=& 50.9915000000.
\end{eqnarray}
As summarised in Table \ref{prediction2}, the predictions for the $p=1$ case are similar to those of $p=2$, in agreement with Planck '15 data with the total number of e-folds $N\approx 888$. In this case, the Hubble parameter during inflation is
\begin{equation}
H_* = \kappa \sqrt{\mathcal V_* / 3} = 5.09 \text{  TeV}.
\end{equation}
Note that for the $p=1$ case, the mass of the gauge boson is $M_{A_\mu}^* = 6.78 \text{ TeV}$, and the mass of the gravitino during inflation is $m^*_{3/2} =  4.72$ TeV, a bit smaller than but comparable with the Hubble scale. 
It would be interesting to investigate how much this would affect the power spectrum and bispectrum, 
inspired by a class of inflation models called quasi-single field inflation \cite{Chen:2009we, Chen:2009zp, Achucarro:2010jv, Baumann:2011nk, Noumi:2012vr, Arkani-Hamed:2015bza,Kahn:2015mla,Hetz:2016ics}, 
which may contain also a light inflaton as well as massive fields of Hubble scale mass, that can amplify non-Gaussianities.  Moreover the $U(1)$ mass is close to the Hubble scale (although slightly above) so that it may produce observable effects in the E-mode polarization spectrum.

\begin{table}
  \centering 
  \begin{tabular}{|c|c|c|}
\hline
$n_s$ & $r$ & $A_s$  \\
\hline
0.959 & $4.143 \times 10^{-22} $ & $2.205  \times 10^{-9} $  \\
\hline
\end{tabular}
  \caption{The theoretical predictions for $p=1$ case with $b=-0.0234$, $c=1 \times 10^{-13}$, $\alpha = -0.781$ and $\xi = 0.3023$. }
  \label{prediction2}
\end{table}

\section{SUGRA spectrum}

The above model can be coupled to MSSM-like fields $\varphi$. 
In this case the multiplet containing the inflaton $\text{Re}(s)$ is considered to be a ``hidden sector'' field which is responsible for breaking supersymmetry by both F- and D-terms, as described above.
The supersymmetry breaking is then communicated to the visible sector (MSSM) via gravity mediation.
We consider the following K\"ahler potential and superpotential.
\begin{align}
 \mathcal K &= \mathcal K(s + \bar s ) + \sum \varphi \bar \varphi, \notag \\
 W &= W_h(s) + W_{\text{MSSM}}, 
\end{align}
where $\mathcal K(s + \bar s)$ is given by eq.~(\ref{Kgeneral}), 
the hidden sector superpotential $W_h(s)$ is given by eq.~(\ref{WfPgeneral}), 
and $W_{\text{MSSM}}$ is the MSSM superpotential, which only depends on the MSSM fields $\varphi$.
The soft supersymmetry breaking terms can be calculated as follows
\begin{align}
 m_0^2 &= e^{\kappa^2 \mathcal K} \left( -2 \kappa^4 W_h(s) \bar W_h(s) + \kappa^2 g^{s \bar s} \left| \nabla_s W_h \right|^2 \right), \notag \\
 A_0 &= \kappa^2 e^{\kappa^2 \mathcal K /2 } g^{s \bar s} K_s \left( \bar W_{\bar s} + \kappa^2 K_s \bar W  \right), \notag \\
 B_0 &= \kappa^2 e^{\kappa^2 \mathcal K /2 } \left( g^{s \bar s} K_s \left( \bar W_{\bar s} + \kappa^2 K_s \bar W  \right) - \bar W \right) .
\end{align}
Here, $m_0^2$ is the scalar soft mass squared. All trilinear couplings are the same and given by $A_0 \hat y_i$, 
where $\hat y_i$ are the Yukawa couplings of the rescaled MSSM superpotential $\hat W_{\text{MSSM}} = e^{\kappa^2 \mathcal K /2} W_{\text{MSSM}}$.
The $B\mu$-term parameter is given by $B_0 \hat \mu$, where $\hat \mu = e^{\mathcal K/2} \mu$.

For $p=2$ the Lagrangian contains a Green-Schwarz term eq.~(\ref{Green-Schwarz}), and the theory is not gauge invariant (without the inclusion of extra fields that are charged under the $U(1)$). 
We therefore focus on $p=1$.
The soft terms can be written in terms of the gravitino mass (see eq.~(\ref{gravitino_mass})) 
\begin{align}
 m_0^2 &=   m_{3/2}^2 \left[ - 2 + \mathcal C \right]  , \notag \\
A_0 &= m_{3/2} \ \mathcal C , \notag  \\ 
B_0 &= A_0 - m_{3/2},
\end{align}
where
\begin{align}
 \mathcal C &= 
 \left. -  \frac{\left( - \xi  e^{\alpha  b^2 \phi^2} +b \phi \left(4 \alpha  \xi  e^{\alpha  b^2 \phi^2}-1\right)+2\right)^2}
{4 \alpha  \xi ^2 e^{2 \alpha  b^2 \phi^2}-4 \alpha  b \xi  \phi e^{\alpha  b^2 \phi^2}+8 \alpha ^2 b^3 \xi  \phi^3 e^{\alpha  b^2 \phi^2}-2} \right|_{\phi = \phi_\text{min} }.
\end{align}
Using the parameters presented in section~\ref{sec:p=1}, we find $m_{3/2} = 18.36$~TeV and $\mathcal C = 1.53$. 
For $\xi = 0$ the model reduces to the one analysed in~\cite{Antoniadis:2015mna}, 
where one has $\mathcal C = 1.52$ and $m_{3/2} = 17.27$~TeV (with $\phi_{\text{min}} = 9.96$). 
As in~\cite{Antoniadis:2015mna}, 
the scalar soft mass is tachyonic.
This can be solved either by introducing an extra Polonyi-like field, or by allowing a non-canonical K\"ahler potential for the MSSM-like fields $\varphi$. 
The resulting low energy spectrum is expected to be similar to the one described in~\cite{Antoniadis:2015mna}.
We do not perform this analysis, but only summarize their results. 

Since the tree-level contribution to the gaugino masses vanishes, their mass is generated at one-loop by the so-called `Anomaly Mediation' contribution~\cite{Randall:1998uk,Giudice:1998xp,Bagger:1999rd}.
As a result, the spectrum consists of 
very light neutralinos ($O(10^2)$ GeV), of which the lightest (a mostly Bino-like neutralino) is the LSP dark matter candidate,
slightly heavier charginos and a gluino in the $1-3$ TeV range.
The squarks are of the order of the gravitino mass ($\sim 10$ TeV), with the exception of the stop squark which can be as light as 2 TeV.

\section{Conclusions}\label{conclusion}
In this paper, we have studied the cosmology of a simple model of supersymmetry breaking, based on a single chiral multiplet (the string dilaton) with a gauged shift symmetry (R-symmetry in an appropriate field basis) leading to a de Sitter vacuum with a tuneable cosmological constant~\cite{Villadoro:2005yq, Knoops}. When coupled to the MSSM, the model have been shown to yield an interesting low energy pattern of supersymmetry breaking, depending on one parameter, distinct from other proposals~\cite{Antoniadis:2015mna, Antoniadis:2015adn}. By modifying the dilaton K\"ahler potential with a non-perturbative correction that could arise from NS5-brane instantons (depending on two extra parameters), we show that the scalar potential can acquire an inflationary plateau producing sufficient inflation, consistent with cosmological observations, without altering the low energy particle spectrum around the minimum.

The predicted tensor-to-scalar ratio is unfortunately too small to be detected. On the other hand, our model may lead to measurable non-gaussianities and/or E-mode polarisation effects that deserve further study.

Our model provides therefore an interesting example of connecting the inflation sector with the `hidden' sector of supersymmetry breaking by identifying the inflaton with the dilaton which is also the scalar partner of the goldstino (sgoldstino). It does not belong however to the class of models studied in the past~\cite{AlvarezGaume:2010rt}, since there is also a D-term component in the order parameter of supersymmetry breaking. It would be interesting to analyse in detail the generalised class of such models.

Another interesting question is the possible implementation of this model in string theory. As mentioned in the introduction, the framework of moduli stabilisation by internal magnetic fields for several abelian factors of the gauge group along the compact directions of the compactified manifold, combined with non-criticality, seems to be a promising direction to explore.

\section*{Acknowledgements}
This work is supported in part by Franco-Thai Cooperation Program in Higher Education and Research `PHE SIAM 2016', in part by the ``CUniverse'' research promotion project by Chulalongkorn University (grant reference CUAASC) and in part by the NCCR SwissMAP funded by the Swiss National Science Foundation. A.C. thanks Phongpichit Channuie for discussion. R.K. would like to thank the CERN Theory Division for its hospitality during part of this work.



\begin{thebibliography}{99}


\bibitem{Villadoro:2005yq}  
G.~Villadoro and F.~Zwirner, ``De-Sitter vacua via consistent D-terms,''  Phys.\ Rev.\ Lett.\  {\bf 95} (2005) 231602  
[hep-th/0508167].
  
\bibitem{Knoops} 
  I.~Antoniadis and R.~Knoops,
  ``Gauge R-symmetry and de Sitter vacua in supergravity and string theory,''
  Nucl.\ Phys.\ B {\bf 886} (2014) 43
  [arXiv:1403.1534 [hep-th]].
  
\bibitem{Ghilencea}
 I.~Antoniadis, D.~M.~Ghilencea and R.~Knoops,
  ``Gauged R-symmetry and its anomalies in 4D N=1 supergravity and phenomenological implications,''
  JHEP {\bf 1502} (2015) 166
  [arXiv:1412.4807 [hep-th]].

  
\bibitem{Antoniadis:2015mna}
  I.~Antoniadis and R.~Knoops,  ``MSSM soft terms from supergravity with gauged R-symmetry in de Sitter vacuum,''
  Nucl.\ Phys.\ B {\bf 902} (2016) 69
  [arXiv:1507.06924 [hep-ph]].
  
\bibitem{Antoniadis:2015adn}
  I.~Antoniadis and R.~Knoops, ``Gauging MSSM global symmetries and SUSY breaking in de Sitter vacuum,''
  Nucl.\ Phys.\ B {\bf 903} (2016) 304
  [arXiv:1511.04283 [hep-ph]].


\bibitem{Antoniadis:1996vw}
  I.~Antoniadis, C.~Bachas, C.~Fabre, H.~Partouche and T.~R.~Taylor,
  ``Aspects of type I - type II - heterotic triality in four-dimensions,''
  Nucl.\ Phys.\ B {\bf 489} (1997) 160
  [hep-th/9608012].
  
\bibitem{Antoniadis:2004pp}
  I.~Antoniadis and T.~Maillard,
  ``Moduli stabilization from magnetic fluxes in type I string theory,''
  Nucl.\ Phys.\ B {\bf 716} (2005) 3
  [hep-th/0412008];
  I.~Antoniadis, A.~Kumar and T.~Maillard,
  ``Magnetic fluxes and moduli stabilization,''
  Nucl.\ Phys.\ B {\bf 767} (2007) 139
  [hep-th/0610246].

\bibitem{Antoniadis:2008uk}
  I.~Antoniadis, J.-P.~Derendinger and T.~Maillard,
  ``Nonlinear N=2 Supersymmetry, Effective Actions and Moduli Stabilization,''
  Nucl.\ Phys.\ B {\bf 808} (2009) 53
  [arXiv:0804.1738 [hep-th]].

\bibitem{AlvarezGaume:2010rt}
  L.~Alvarez-Gaume, C.~Gomez and R.~Jimenez,
  ``Minimal Inflation,''
  Phys.\ Lett.\ B {\bf 690} (2010) 68
  [arXiv:1001.0010 [hep-th]];
  ``A Minimal Inflation Scenario,''
  JCAP {\bf 1103} (2011) 027
  [arXiv:1101.4948 [hep-th]].

 \bibitem{Dine:2011ws} 
  M.~Dine and L.~Pack,
  ``Studies in Small Field Inflation,''
  JCAP {\bf 1206}, 033 (2012)
  [arXiv:1109.2079 [hep-ph]];
  A.~Achucarro, S.~Mooij, P.~Ortiz and M.~Postma,
  ``Sgoldstino inflation,''
  JCAP {\bf 1208} (2012) 013
  [arXiv:1203.1907 [hep-th]];
  S.~Ferrara and D.~Roest,
  ``General sGoldstino Inflation,''
  JCAP {\bf 1610} (2016) no.10,  038
  [arXiv:1608.03709 [hep-th]].
  

\bibitem{Freedman:2012zz}
  D.~Z.~Freedman and A.~Van Proeyen, ``Supergravity,''  Cambridge, UK: Cambridge Univ. Pr. (2012) 607 p

 \bibitem{Kawasaki:2000yn}
   M.~Kawasaki, M.~Yamaguchi and T.~Yanagida,
   ``Natural chaotic inflation in supergravity,''
   Phys.\ Rev.\ Lett.\  {\bf 85}, 3572 (2000)
   [hep-ph/0004243];
   A.~Mazumdar, T.~Noumi and M.~Yamaguchi,
   ``Dynamical breaking of shift-symmetry in supergravity-based inflation,''
   Phys.\ Rev.\ D {\bf 90}, no. 4, 043519 (2014)
   [arXiv:1405.3959 [hep-th]].
  
  
\bibitem{Ade:2015lrj} P.~A.~R.~Ade {\it et al.} [Planck Collaboration], ``Planck 2015 results. XX. Constraints on inflation,'' arXiv:1502.02114 [astro-ph.CO].



\bibitem{Chen:2009we}
  X.~Chen and Y.~Wang,
  ``Large non-Gaussianities with Intermediate Shapes from Quasi-Single Field Inflation,''
  Phys.\ Rev.\ D {\bf 81} (2010) 063511
  [arXiv:0909.0496 [astro-ph.CO]].
  
\bibitem{Chen:2009zp}
  X.~Chen and Y.~Wang,
  ``Quasi-Single Field Inflation and Non-Gaussianities,''
  JCAP {\bf 1004} (2010) 027
  [arXiv:0911.3380 [hep-th]].

\bibitem{Achucarro:2010jv}
  A.~Achucarro, J.~O.~Gong, S.~Hardeman, G.~A.~Palma and S.~P.~Patil,
  ``Mass hierarchies and non-decoupling in multi-scalar field dynamics,''
  Phys.\ Rev.\ D {\bf 84} (2011) 043502
  [arXiv:1005.3848 [hep-th]];  
   A.~Achucarro, J.~O.~Gong, S.~Hardeman, G.~A.~Palma and S.~P.~Patil,
  JHEP {\bf 1205} (2012) 066
  [arXiv:1201.6342 [hep-th]].
  
\bibitem{Baumann:2011nk}
  D.~Baumann and D.~Green,
  ``Signatures of Supersymmetry from the Early Universe,''
  Phys.\ Rev.\ D {\bf 85} (2012) 103520
  [arXiv:1109.0292 [hep-th]].

 \bibitem{Noumi:2012vr}
   T.~Noumi, M.~Yamaguchi and D.~Yokoyama,
   ``Effective field theory approach to quasi-single field inflation and effects of heavy fields,''
   JHEP {\bf 1306}, 051 (2013)
   [arXiv:1211.1624 [hep-th]].
 
\bibitem{Arkani-Hamed:2015bza}
  N.~Arkani-Hamed and J.~Maldacena, ``Cosmological Collider Physics,'' arXiv:1503.08043 [hep-th].

\bibitem{Kahn:2015mla}
  Y.~Kahn, D.~A.~Roberts and J.~Thaler,
``The goldstone and goldstino of supersymmetric inflation,''
  JHEP {\bf 1510} (2015) 001
  [arXiv:1504.05958 [hep-th]].

\bibitem{Hetz:2016ics}
  A.~Hetz and G.~A.~Palma,
  ``Sound Speed of Primordial Fluctuations in Supergravity Inflation,''
  Phys.\ Rev.\ Lett.\  {\bf 117} (2016) no.10,  101301
  [arXiv:1601.05457 [hep-th]].
  
\bibitem{Randall:1998uk}  L.~Randall and R.~Sundrum, ``Out of this world supersymmetry breaking,''  Nucl.\ Phys.\ B {\bf 557} (1999) 79   
[hep-th/9810155].
\bibitem{Giudice:1998xp}   G.~F.~Giudice, M.~A.~Luty, H.~Murayama and R.~Rattazzi, ``Gaugino mass without singlets,''  JHEP {\bf 9812} (1998) 027  
[hep-ph/9810442].   
\bibitem{Bagger:1999rd}   J.~A.~Bagger, T.~Moroi and E.~Poppitz, ``Anomaly mediation in supergravity theories,''  JHEP {\bf 0004} (2000) 009  
[hep-th/9911029].
  
 


\end{thebibliography}
\end{document}